\documentstyle[11pt]{article}
\begin{document}
\begin{large}
\begin{titlepage}

\vspace{0.2cm}

\title{Pair production of neutralinos via photon-photon collisions
\footnote{The project supported by National Natural Science
          Foundation of China}}
\author{{Zhou Fei$^{b}$, Ma Wen-Gan$^{a,b}$, Jiang Yi$^{b}$ and Han Liang$^{b}$}\\
{\small $^{a}$CCAST (World Laboratory), P.O.Box 8730, Beijing 100080,
China.}\\
{\small $^{b}$Department of Modern Physics, University of Science
        and Technology}\\
{\small of China (USTC), Hefei, Anhui 230027, China.}  }
\date{}
\maketitle

\vskip 12mm

\begin{center}\begin{minipage}{5in}

\vskip 5mm
\begin{center} {\bf Abstract}\end{center}
\baselineskip 0.3in
{We investigated the production of neutralino pairs via photon-photon
collisions in the minimal supersymmetric model(MSSM) at future
linear colliders. The numerical analysis of their production rates is
carried out in the mSUGRA scenario. The results show that this cross
section can reach about 18 femto barn for $\tilde{\chi}^{0}_{1}
\tilde{\chi}^{0}_{2}$ pair production and 9 femto barn for
$\tilde{\chi}^{0}_{2}\tilde{\chi}^{0}_{2}$ pair production with our
chosen input parameters. }\\

\vskip 5mm
{PACS number(s): 14.80.Ly, 12.15.Ji, 12.60.Jv}
\end{minipage}
\end{center}
\end{titlepage}

\baselineskip=0.36in

\eject
\rm
\baselineskip=0.36in

\begin{flushleft} {\bf 1. Introduction} \end{flushleft}
In the supersymmetric theory\cite{Nilles}, proper electroweak symmetry
breaking induces the right properties of the lightest supersymmetric
particle (LSP) to be a natural candidate for weak-interacting cold dark matter,
which can explain many astrophysical observations\cite{darkm}.
The minimal supersymmetric standard model(MSSM)\cite{Nilles} predicts that
there exists an absolutely stable LSP.
Most often the LSP in the MSSM theory is the lightest Majorana fermionic
neutralino $\tilde{\chi}^{0}_{1}$. Therefore the production of the lightest
neutralino $\tilde{\chi}^{0}_{1}$ and the second lightest neutralino
$\tilde{\chi}^{0}_{2}$ may be studied at present and future experiments
and the detailed study of the neutralino sector will help us to
determine which kind of the supersymmetric models really exists in nature.
In the MSSM, the physical neutralino mass eigenstates $\tilde{\chi}^{0}_{i}~~
(i=1 \sim 4)$ are the combinations of the neutral gauginos($\tilde{B}$,
$\tilde{W}^{3}$) and the neutral higgsinos($\tilde{H}^{0}_{1}$,
$\tilde{H}^{0}_{2}$). In the two-component fermion
fields $\psi^{0}_{j}= (-i\lambda^{'}, -i\lambda^{3}, \psi_{H^{0}_{1}},
\psi_{H^{0}_{2}})$\cite{haber}, where $\lambda^{'}$ is the bino and
$\lambda^{3}$ is the neutral wino, the neutralino mass matrix in the
Lagrangian is given by
$$
{\cal L}_{M}=-\frac{1}{2}(\psi^{0})^{T} Y \psi^{0} +h.c.,
\eqno{(1.1)}
$$
where the matrix $Y$ reads
$$
Y=\left (
\begin{array}{cccc}
M_{1} & 0& -m_{Z}\sin\theta_{W}\cos\beta & m_{Z}\sin\theta_{W}\sin\beta\\
0& M_{2} & m_{Z}\cos\theta_{W}\cos\beta & -m_{Z}\cos\theta_{W}\sin\beta\\
-m_{Z}\sin\theta_{W}\cos\beta & m_{Z}\cos\theta_{W}\cos\beta& 0 & -\mu\\
m_{Z}\sin\theta_{W}\sin\beta & -m_{Z}\cos\theta_{W}\sin\beta &-\mu & 0
\end{array}\right)
\eqno{(1.2)}
$$
In the above equation, the neutralino mass matrix is related to four unknown
parameters, namely $\mu$, $M_2$, $M_1$ and $\tan\beta=v_2/v_1$, ratio
of the vacuum expectation values of the two Higgs fields. $\mu$ is the
supersymmetric Higgs-boson-mass parameter and $M_2$ and $M_1$ are the
gaugino mass parameters associated with the $SU(2)$ and $U(1)$ subgroups,
respectively. In CP-noninvariant theories, $M_{1}$, $M_{2}$
and higgsino mass parameter $\mu$ can be complex. However,
$M_{2}$ can be real and positive without loss of generality by
reparametrization of the fields. In this work we shall investigate
neutralino pair production in framework of the MSSM while ignoring CP-violation
and taking $M_1$ and $\mu$ as being real. The matrix $Y$ is symmetric and can
be diagonalized by one unitary matrix $N$ such that $N_{D}=N^{*}YN^{+}=
diag(m_{\tilde{\chi}^{0}_{1}}, m_{\tilde{\chi}^{0}_{2}},
m_{\tilde{\chi}^{0}_{3}},m_{\tilde{\chi}^{0}_{4}})$
in order of $m_{\tilde{\chi}^{0}_{1}}\le m_{\tilde{\chi}^{0}_{2}}\le
m_{\tilde{\chi}^{0}_{3}}\le m_{\tilde{\chi}^{0}_{4}}$. Then the two-component
mass eigenstates can be
$$
\chi^{0}_{i}=N_{ij}\psi^{0}_{j},~~~i, j=1,\dots,4.
\eqno{(1.3)}
$$
The proper four-component mass eigenstates are defined in terms of
two-component fields as
$$
\tilde{\chi}^{0}_{i}=\left(\begin{array}{c}
\chi^{0}_{i}\\
\bar{\chi}^{0}_{i}\end{array}\right)
~~~(i=1,\dots, 4),
\eqno{(1.4)}
$$
and the mass term becomes
$$
{\cal L}_{m}=-\frac{1}{2}\sum_{i}\tilde{M}_{i} \bar{\tilde{\chi}^{0}_{i}}
\tilde{\chi}^{0}_{i},
\eqno{(1.5)}
$$
where $\tilde{M}_{i}$'s are the diagonal elements of $N_{D}$.
\par
The future $e^{+} e^{-}$ Linear Colliders(LC) are designed to give
facilities for $e^+ e^-$, $\gamma \gamma$ and other collisions at
the energy of $500 \sim 2000~GeV$ with a
luminosity of the order $10^{33} cm^{-2} s^{-1}$ \cite{a1}.
For example, the proposed TESLA collider is known as a powerful tool for
exploration the multi-handred GeV scale\cite{brin}. Different options of
this machine, namely, $e^+e^-$, $\gamma\gamma$, $\gamma e$ and $e^-e^-$
are complementary to each other and will add essential new information
to that obtained from the LHC. Its annual ($10^7$ s) $\gamma\gamma$ luminosity
will be about $10-30~fb^{-1}$ (in the high energy peak) with possible
upgrade of luminosity by one order of magnitude.
Searching for supersymmetric particles and determining their properties
are one of the main tasks at future LC. In detecting the existence of
neutralinos, both $e^{+}e^{-}$ and $\gamma \gamma$ collisions have clearer
background than hardron collisions, but $\gamma\gamma$ collision at LC would
have distinct advantage over the situation of LC operating
in the $e^{+}e^{-}$ colision mode, where the resonant effects of Higgs
bosons can be observed only at some specific center of mass energy ranges
of the machine and the possible s-channel suppression would generally
reduce the cross section of neutralino pair production. Because of the
continuous c.m.s energy distribution of the colliding photons back-scattered
by $e^{-}$ and $e^{+}$ beams, the intermediate resonant
effects of Higgs bosons could enhance the neutralino pair production rate
over a rather wide colliding energy range at electron-positron colliders.
\par
So far there is no experimental
evidence for neutralinos at CERN LEP2. They only set lower bound on the
lightest neutralino mass $m_{\tilde{\chi}_{1}^{0}}$. Recent experimental
reports presented that the mass of the lightest neutralino may be larger than
$32.5~GeV$\cite{L3} and the lower limit of the chargino mass is $76.8~GeV$
\cite{L3a}. The direct neutralino pair productions at the CERN Large Hadron
Collider (LHC) are studied in references \cite{beenakker}\cite{jiang}\cite{han}.
The production of neutralino pair can be produced also at the LC machine
operating in both $e^- e^+$ and $\gamma \gamma$ collision modes. Recently,
it has been found that the $\tilde{\chi}^{0}_{1}\tilde{\chi}^{0}_{2}$
production rate in $e^+ e^-$ collision mode can reach several handred
femto barn\cite{a2}.
\par
In this paper we investigate the potential of direct neutralino pair
production at the LC operating in photon-photon collision mode in framework
of the MSSM with complete one-loop Feynman diagrams. The numerical calculation
will be illustrated in the CP-conserving mSUGRA scenario with five input
parameters, namely $m_{1/2}$, $m_0$, $A_0$, $\mu$ and $\tan\beta$, where
$m_{1/2}$, $m_0$ and $A_0$ are the universal gaugino mass, scalar mass at
GUT scale and the trilinear soft breaking parameter in the superpotential
respectively. From these five parameters, all the masses and couplings of
the model are determined by evolution from the GUT scale down to the
low electroweak scale \cite{msugra}. The paper is organized as follows: In
section 2, we give calculations of the neutralino pair production at the LC
operating in photon-photon collision mode. In section 3, we discuss the
numerical results of the cross sections. A short summary is presented in
section 4. Finally the explicit expressions of form factors for s-channel
diagrams are listed.

\begin{flushleft} {\bf 2. The Calculation of $e^{+}e^{-} \rightarrow
    \gamma\gamma \rightarrow \tilde{\chi}^{0}_{i}\tilde{\chi}^{0}_{j}$}
\end{flushleft}
\par
Neutralinos can be produced at $e^+e^-$ colliders, either in diagonal or
in mixed pairs. In this section we calculate the processes
$$
e^+e^- \rightarrow \gamma\gamma \rightarrow
       \tilde{\chi}_i^{0} \tilde{\chi}_j^{0},~~(i=1,2,~j=2).
$$
The generic Feynman diagrams contributing to the subprocess
$\gamma\gamma \rightarrow \tilde{\chi}_i^{0} \tilde{\chi}_j^{0}$ in
the MSSM at one-loop level are depicted in Fig.1, where the exchange of
incoming photons in Fig.1$(a.1\sim6)$, Fig.1(b.5) and Fig.1$(c.1\sim4)$
are not shown. Fig.1($a.1 \sim 6$) are box diagrams. Fig.1($b.1 \sim 5$) are
quartic interaction diagrams. Fig.1($c.1 \sim 3$) represent triangle diagrams.
The Fig.1$(b.3 \sim b.5)$, $(c.1\sim 3)$ are also called s-channel diagrams.
The Feynman diagrams in Fig.1 include the loops of quarks(U,D),
squarks $(\tilde{U},\tilde{D})$, leptons(E), sleptons$(\tilde{E})$ of three
generations, charginos, $W^{+}$ boson, charged ghost particles, charged Higgs
boson and goldstone. Due to the Yukawa couplings strength, the contributions
from the loop diagrams of the third generation quarks and squarks are more
important than those from other diagrams. Here we should mention two points:
(1) There is no diagram with a triangle squark(slepton) loop
coupling with an $A^0$ or $G^0$ boson in Fig.1, because the vertices of
$A^{0}(G^{0})-\tilde{q}-\tilde{q}$ vanish\cite{hunter}. (2) Our
calculation shows that the Feynman diagrams involving quartic vertices
$\gamma-h^0-G^+(H^+)-G^-(H^-)$ and $\gamma-H^0-G^+(H^+)-G^-(H^-)$, which have
similar structure to Fig.1(b.5), have no contribution to the cross section,
therefore omitted in figure 1.
The $Z^0$ boson intermediated s-channel diagrams similar to the diagrams of
Fig.1(b.3) $\sim$ (b.5), Fig.1(c.1) $\sim$ (c.3) are not plotted in Fig.1
either, since they cannot contribute to the cross section. For this result
there are two reasons: (1) The CP-odd scalar component of the $Z^0$ boson
does not couple to the invariant CP-even $\tilde{\chi}_i^{0} \tilde{\chi}_j^{0}$
state. (2) The vector component of the $Z^0$ boson wave function does not
couple to the initial $\gamma\gamma$ state as the result of the Laudau-Yang
theorem.
\par
Since there is no tree level diagram for the neutralino pair production
via photon-photon collisions, the calculation for this process can be
simply carried out by summing all unrenormalized reducible and irreducible
one-loop diagrams and the results will be finite and gauge invariant. In
this work, we perform the calculation in the 't Hooft-Feynman gauge, and
take the CKM matrix as identity.
\par
We denote the reaction of neutralino pair production via photon-photon
collisions as:
$$
\gamma (p_1, \mu) \gamma (p_2, \nu) \longrightarrow
\tilde{\chi}_{i}^{0} (k_1) \tilde{\chi}_{j}^{0} (k_2),~~(i=1,2,~j=2).
 \eqno{(2.1)}
$$
where $p_1,~p_2$ and $k_1,~k_2$ denote the four momenta of the incoming
photons and outgoing neutralinos, respectively. In calculation of the
amplitude, one should note that there should be a relative sign
$(-1)^{\delta_{ij}}$ between the amplitudes of one diagram and its
counterpart obtained by exchanging the final neutralinos as a result
of Fermi statistics. The corresponding matrix element can be written as
\begin{eqnarray*}
{\cal M}&=&{\cal M}_{s}+{\cal M}_{b}+{\cal M}_{q}= {\epsilon^{\mu}(p_1)
\epsilon^{\nu}(p_2) \overline{u}(k_2)}
P_{L} \left\{ f_{1}^{L}\gamma_{\mu}\gamma_{\nu}+
f_{2}^{L}\gamma_{\mu}\gamma_{\nu}\rlap/{p}_1+
f_{3}^{L}\gamma_{\nu}\gamma_{\mu}+
f_{4}^{L}\gamma_{\nu}\gamma_{\mu}\rlap/{p}_1\right. \\
&+&f_{5}^{L} p_{2\mu} \gamma_{\nu} 
\left. +f_{6}^{L} p_{2\mu} \gamma_{\nu} \rlap/{p}_1 +
f_{7}^{L} k_{1\mu} \gamma_{\nu}+
f_{8}^{L} k_{1\mu} \gamma_{\nu} \rlap/{p}_1+
f_{9}^{L} p_{1\nu} \gamma_{\mu}+
f_{10}^{L} p_{1\nu} \gamma_{\mu} \rlap/{p}_1 +
f_{11}^{L} k_{1\nu} \gamma_{\mu} \right. \\
&+&\left. f_{12}^{L} k_{1\nu} \gamma_{\mu} \rlap/{p}_1 +
f_{13}^{L} p_{2\mu} p_{1\nu} +
f_{14}^{L} p_{2\mu} p_{1\nu} \rlap/{p}_1 +
f_{15}^{L} p_{2\mu} k_{1\nu} +
f_{16}^{L} p_{2\mu} k_{1\nu} \rlap/{p}_1 +
f_{17}^{L} k_{1\mu} k_{1\nu} \right. \\
&+&\left. f_{18}^{L} k_{1\mu} k_{1\nu} \rlap/{p}_1 +
f_{19}^{L} k_{1\mu} p_{1\nu} +
f_{20}^{L} k_{1\mu} p_{1\nu} \rlap/{p}_1 +
\epsilon_{\mu \nu \alpha \beta} \left[
f_{21}^{L} {p_1}^{\alpha} {p_2}^{\beta}+
f_{22}^{L} {p_1}^{\alpha} {p_2}^{\beta} \rlap/{p}_1 \right] \right\} v(k_1) \\
&+& (P_L \rightarrow P_R, f_{i}^{L} \rightarrow f_{i}^{R}).
~~~~~~~~~~~~~~~~~~~~~~~~~~~~~~~~~~~~~~~~~~~(2.2)
\end{eqnarray*}
where $P_{L,R}=\frac{1}{2}(1\mp\gamma_5)$, $f_{k}^{L,R}(k=1,...,22)$ are form
factors. ${\cal M}_{s}$ ,${\cal M}_{b}$ and ${\cal M}_{q}$ are the amplitudes of
the s-channel, box and quartic interaction diagrams, respectively. Since our 
calculation shows that the
contribution to the cross section is predominantly from s-channel diagrams,
we listed the explicit expressions of the form factors for s-channel diagrams
in Appendix and omitted the form factor expressions of box and quartic interaction 
diagrams, because
they have long-winded expressions and negligible contribution to cross section
in the resonant effect energy region. Then the cross section for this
subprocess at one-loop order in unpolarized photon collisions can be
obtained by
$$
 \hat{\sigma}(\hat{s},\gamma\gamma \rightarrow \tilde{\chi}_{i}^{0}
 \tilde{\chi}_{j}^{0}) = \frac{1}{16 \pi \hat{s}^2}(\frac{1}{2})^{\delta_{ij}}
             \int_{\hat{t}^{-}}^{\hat{t}^{+}} d\hat{t}~
             \bar{\sum\limits_{}^{}} |{\cal M}|^2,~~~~(i=1,~j=1,2).
\eqno{(2.3)}
$$
In the above equation,
$\hat{t}^\pm=1/2\left[ (m^{2}_{\tilde{\chi}_{i}}+m^{2}_{\tilde{\chi}_{j}}
-\hat{s})\pm \sqrt{(m^{2}_{\tilde{\chi}_{i}}+m^{2}_{\tilde{\chi}_{j}}-\hat{s})^2
-4m^{2}_{\tilde{\chi}_{i}}m^{2}_{\tilde{\chi}_{j}}}\right]$.
The factor $(\frac{1}{2})^{\delta_{ij}}$ is due to the two identical particles
in the final states. The bar over the sum means average over initial spins.
\par
The neutralino pair production via photon-photon fusion
is only a subprocess of the parent $e^{+}e^{-}$ linear collider.
The total cross section of the neutralino pair production via
photon fusion in $e^{+}e^{-}$ collider can be obtained by folding the cross
section of the subprocess $\hat{\sigma} (\gamma\gamma \rightarrow
\tilde{\chi}^{0}_i {\tilde{\chi}}^{0}_j)$ with the photon luminosity.
$$
\sigma(s)=\int_{(m_{\tilde{\chi}_{i}}+m_{\tilde{\chi}_{j}})/
\sqrt{s} }^{x_{max}} dz \frac{dL_{\gamma\gamma}}{dz} \hat{\sigma}
(\gamma\gamma\rightarrow \tilde{\chi}^{0}_{i} \tilde{\chi}^{0}_{j}
\hskip 3mm at \hskip 3mm \hat{s}=z^2 s),
\eqno{(2.5)}
$$
where $\sqrt{s}$ and $\sqrt{\hat{s}}$ are the $e^{+}e^{-}$ and $\gamma \gamma$
c.m.s. energies respectively and $d{\cal L}_{\gamma\gamma}/dz$
is the distribution function of photon luminosity, which is
$$
\frac{d{\cal L}_{\gamma\gamma}}{dz}=2z\int_{z^2/x_{max}}^{x_{max}}
 \frac{dx}{x} f_{\gamma/e}(x)f_{\gamma/e}(z^2/x),
\eqno{(2.6)}
$$
where $f_{\gamma/e}$ is the photon structure function of the electron beam
\cite{sd,sh}. We take the structure function of the photon produced by
Compton backscattering as \cite{sd,si}
\[
f^{Comp}_{\gamma/e}(x)=\left\{
\begin{array}{cl}
\frac{1}{D(\xi)} \left(1-x+\frac{1}{1-x}-\frac{4x}{\xi(1-x)}+\frac{4 x^2}{{\xi}^2 (1-x)^2} \right),
& {\rm for}~x~<~0.83,\\
0, & {\rm for}~x~>~0.83,
\end{array}
\right.
\]
$$
\eqno{(2.7)}
$$
where
$$
D(\xi)=(1-\frac{4}{\xi}-\frac{8}{{\xi}^2})\ln{(1+\xi)}+\frac{1}{2}+\frac{8}{\xi}-\frac{1}{2{(1+\xi)}^2},
~~~~\xi=\frac{4E_0 \omega_0}{{m_e}^2}.\eqno{(2.8)}
$$
Taking $\omega_0$ the maximal energy of backscattering photons, or,
${\xi}=2(1+\sqrt{2})$, we have $D(\xi)=1.8397$.

\par
\begin{flushleft} {\bf 3. Numerical results and discussions} \end{flushleft}
\par
In this section, we present some numerical results of the total cross
section in the mSUGRA scenario from the complete one-loop diagrams for the
process $e^{+}e^{-} \rightarrow \gamma \gamma \rightarrow \tilde{\chi}^{0}_{i}
\tilde{\chi}^{0}_{j},~(i=1,~j=1,2)$. The input parameters are chosen as:
$m_t=175~GeV$, $m_{Z}=91.187~GeV$, $m_b=4.5~GeV$, $\sin^2{\theta_{W}}=0.2315$,
and $\alpha = 1/128$. We assume that $\tilde{\chi}^{0}_{1}$ is the LSP and
escapes detection. In numerical calculation to get the low energy scenario
from the mSUGRA, the renormalization group equations(RGE's)\cite{RGE} are
run from the weak scale $M_Z$ up to the GUT scale, taking all thresholds into
account. We use two loop RGE's only for the gauge couplings and the one-loop
RGE's for the other supersymmetric parameters. The GUT scale boundary
conditions are imposed and the RGE's are run back to $M_Z$, again taking
threshold into account.
\par
As we know that the s-channel resonance effects of the intermediate Higgs
bosons (see Fig.1(b,c)) could play an important role in some c.m.s. energy
regions of incoming photons. With the mSUGRA scenario input parameters used
in our numerical calculation, the mass of the lightest Higgs boson $h^0$ is
under the thresholds of $W^+W^-$ and $Z^0Z^0$ decays. Therefore the relevant
decay width of $h^0$ is mainly contributed by the decay channel of
$h^0 \rightarrow b\bar{b}$. The decay channels of $H^0$ may involve
$H^0 \rightarrow q\bar{q}$(where q may be top and bottom quarks),
$H^0 \rightarrow h^0h^0~(A^0A^0)$, $H^0 \rightarrow \tilde{\chi}^{+}_{1}
\tilde{\chi}^{-}_{1}~ (\tilde{\chi}^{0}_{i}\tilde{\chi}^{0}_{j}(i=1,~j=1,2))$
and $H^0 \rightarrow W^+W^-~(Z^0Z^0)$, if the mass of $H^0$ is larger than
the thresholds of all those decay channels. The main decay channels for the
pseudo-scalar Higgs boson are $A^0 \rightarrow Z^0 h^0$,
$A^0 \rightarrow q\bar{q}$ and $A^0 \rightarrow \tilde{\chi}^{+}_{1}
\tilde{\chi}^{-}_{1}~ (\tilde{\chi}^{0}_{i}\tilde{\chi}^{0}_{j}(i=1,~j=1,2))$
under similar conditions. In our calculation, all the decay widths of the
intermediate Higgs bosons are considered at the tree level and their
expressions can be found in appendix B of reference \cite{hunter}.
\par
The cross sections for $\tilde{\chi}^{0}_{1}\tilde{\chi}^{0}_{2}$
and $\tilde{\chi}^{0}_{2}\tilde{\chi}^{0}_{2}$ via photon collisions at linear
colliders versus the mass of $\tilde{\chi}_{2}^{0}$ is shown in Fig.2.
The input parameters are chosen as $m_{0}=100~GeV$, $A_{0}= 300~GeV$, $\mu>0$
and $\tan{\beta}=4$. We calculate the cross sections at the collision energies
of electron-positron $\sqrt{s}$ being $500~GeV$ and $1~TeV$, respectively.
In framework of the mSUGRA, when the mass of $\tilde{\chi}^{0}_{2}$
varies from $81~GeV$ to $251~GeV$ as shown in Fig.2, chargino mass
has the values above the lower limit give in Ref.\cite{L3a}, the masses
of Higgs boson $H^0$ and $A^0$ increase from $231~GeV$ to $510~GeV$ and
from $226~GeV$ to $507~GeV$ respectively and the mass of the Higgs boson
$h^0$ is always under the threshold of $\tilde{\chi}^{0}_{1}
\tilde{\chi}^{0}_{2}$ decay. Therefore the contribution to
the cross section from the s-channel with exchanging $h^0$ is very small
due to the s-channel suppression. The masses of $H^0$ and $A^0$ may reach
the thresholds of $\tilde{\chi}^{0}_{1} \tilde{\chi}^{0}_{2}$ and
$\tilde{\chi}^{0}_{2} \tilde{\chi}^{0}_{2}$ decay modes, then the cross
section of the subprocess will be strongly enhanced by the
s-channel resonant effects when the total energy $\sqrt{\hat{s}}$ of the
subprocess approaches the mass of $H^0$ or $A^0$.
\par
We can see from Fig.2 that all the cross sections for the neutralino
pairs $\tilde{\chi}^{0}_{i} \tilde{\chi}^{0}_{j}~(i=1,2,j=1)$ decreases
with the increment of the value of $m_{\tilde{\chi}_2}$. It can reach 18
femto barn when $\sqrt{s}=500~GeV$ and $m_{\tilde{\chi}_{2}} \sim 81~GeV$.
For the two curves of $\sqrt{s}=500~GeV$, the cross sections with
$\sqrt{s}=500~GeV$ go down rapidly when the values of the mass of
$\tilde{\chi}^{0}_{2}$ are in the vicinity of $193~GeV$. The reason is that the
resonant effects from $H^0$ and $A^0$ exchanging s-channels do not contribute
to the total cross section, since the maximum c.m.s energy of
incoming photons is $\sqrt{\hat{s}_{max}}\sim 0.83 \sqrt{s}<m_{H^0}(m_{A^0})$.
That fact reflects that the resonant effect of the s-channels plays an
important role in the total cross sections of neutralino pair productions at LC.
From this figure we can see that the cross sections of neutralino pair
productions with $\sqrt{s}=500~GeV$ are always larger than those with
$\sqrt{s}=1~TeV$ when $m_{\tilde{\chi}^{0}_{2}}$ is less than $180~GeV$, and
will be smaller than those with $\sqrt{s}=1~TeV$ when $m_{\tilde{\chi}_{2}}$
is larger than $185~GeV$.
\par
In Fig.3 we present the cross sections of neutralino pair productions
versus $\tan{\beta}$ with the colliding energy of electron and positron being
$500~GeV$ and $1~TeV$ respectively, where the other four input parameters
are chosen as $m_{0}=100~GeV$, $m_{1/2}=150~GeV$, $A_{0}= 300~GeV$ and
$\mu>0$. Our calculation in the mSUGRA model shows that when
$\tan\beta$ increases from 2 to 32, the mass of $\tilde{\chi}^{0}_{1}$
ranges from $51~GeV$ to $57~GeV$ and the mass of $\tilde{\chi}^{0}_{2}$
ranges from $98~GeV$ to $102~GeV$. So the dependence of the masses of
neutralinos on $\tan\beta$ is very weak, but the masses of Higgs boson
$H^0$ and $A^0$ depend on $\tan\beta$ obviously and decrease from $352~GeV$
to $166~GeV$ and from $344~GeV$ to $165~GeV$, respectively. The mass of
$h^0$ is a function of $\tan\beta$ too, but keeps
$m_{h^0}<m_{\tilde{\chi}^{0}_{1}} + m_{\tilde{\chi}^{0}_{2}}$. Since the
couplings of Higgs bosons to quarks and squarks pair of the third generation
are related to the ratio of the vacuum expectation values, $\tan{\beta}$
should effect the cross sections substantially due to Yukawa coupling strength.
The line shapes in this figure are determined mainly by parts of the form
factors of $f_{1,s}^{L,R}$, $f_{13,s}^{L,R}$ and $f_{21,s}^{L,R}$, which come
from the diagrams involving couplings between heavy quarks(top or bottom quark)
and Higgs bosons. The form factors from these diagrams are all proportional
to a factor, which has the form as $\frac{A}{\cos{\beta}}+
\frac{B}{\sin{\beta}}$. Therefore all the
curves in Figure 3 have similar line shapes. We can see from this figure that
the cross section for mixed neutralino pair
($\tilde{\chi}^{0}_{1}\tilde{\chi}^{0}_{2}$) production is
larger than that for diagonal pair($\tilde{\chi}^{0}_{2}\tilde{\chi}^{0}_{2}$)
production. When the ratio of both vacuum expectation values $\tan{\beta}$ is
in the vicinity of $\tan{\beta} \sim 5$, the cross sections for both
$\tilde{\chi}^{0}_{1} \tilde{\chi}^{0}_{2}$ and $\tilde{\chi}^{0}_{2}
\tilde{\chi}^{0}_{2}$ pair productions can reach maximum, where we have
$m_{\tilde{\chi}_{1}}= 52~GeV$ and $m_{\tilde{\chi}_{2}}= 98~GeV$.

\par
\begin{flushleft} {\bf 4. Summary} \end{flushleft}
\par
In this paper, we studied the pair production processes of the neutralinos
via photon-photon fusion at LC. The numerical analysis of their production
rates is carried out in the mSUGRA scenario with some typical parameter sets.
The results show that the cross section of the neutralino pair productions
via photon-photon collisions can reach about 18 femto barn for
$\tilde{\chi}^{0}_{1}\tilde{\chi}^{0}_{2}$ pair production and 9 femto barn
for $\tilde{\chi}^{0}_{2}\tilde{\chi}^{0}_{2}$ pair production at an
electro-positron LC operating in $\gamma \gamma$ collision mode with our
chosen parameters, which is one order smaller quantitatively than
at the machine operating in $e^+ e^-$ collision mode. At future TESLA collider,
the annual $\gamma\gamma$ luminosity is designed to be $10-30~{fb}^{-1}$ and
one order higher after upgraded\cite{brin}, these translate into about
90 $\sim$ 270 events per year (about $10^3$ events after upgrade) for
$\tilde{\chi}^{0}_{2}\tilde{\chi}^{0}_{2}$ production and 180 $\sim$ 540 events
per year (few thausand events after upgrad) for $\tilde{\chi}^{0}_{1}
\tilde{\chi}^{0}_{2}$ production. Our numerical results present a fact that
in some c.m.s energy regions of incoming photons, there exist obvious
resonance effects from s-channel diagrams by exchanging intermediate
$H^0$ and $A^0$ Higgs bosons, which can makes observable
enhancement on the cross sections of the parent processes at
linear colliders.

\vskip 4mm
\noindent{\large\bf Acknowledgement:}
This work was supported in part by the National Natural Science
Foundation of China(project number: 19875049), the Youth Science
Foundation of the University of Science and Technology of China, a grant
from the Education Ministry of China and the State Commission of Science
and Technology of China.

\vskip 5mm
\begin{center} {\Large Appendix}\end{center}
\par
In this appendix we list the form factors for the one-loop s-channel diagrams.
The amplitude of s-channel diagrams, which was defined in Eq.(2.2), has
the form
\begin{eqnarray*}
{\cal M}_{s}&=& \frac{e^2}{16 {\pi}^2}{\epsilon^{\mu}(p_1)
\epsilon^{\nu}(p_2) \overline{u}(k_2)}
P_{L} \left\{ f_{1,s}^{L}\gamma_{\mu}\gamma_{\nu}+
f_{3,s}^{L}\gamma_{\nu}\gamma_{\mu}
+ f_{13,s}^{L} p_{2\mu} p_{1\nu} +
f_{21,s}^{L} \epsilon_{\mu \nu p_1 p_2} \right\} v(k_1) \\
&+& (P_L \rightarrow P_R, f_{k,s}^{L} \rightarrow f_{k,s}^{R})\\
&=&\frac{e^2}{16 {\pi}^2}{\epsilon^{\mu}(p_1) \epsilon^{\nu}(p_2) \overline{u}(k_2)}
P_{L} \left\{ 2f_{1,s}^{L}g_{\mu\nu}
+ f_{13,s}^{L} p_{2\mu} p_{1\nu} +
f_{21,s}^{L} \epsilon_{\mu \nu p_1 p_2} \right\} v(k_1) \\
&+& (P_L \rightarrow P_R, f_{k,s}^{L} \rightarrow f_{k,s}^{R}),
\end{eqnarray*}
where $f_{1,s}^{L}=f_{3,s}^{L}$, $f_{1,s}^{R}=f_{3,s}^{R}$.
\par
We divide the form factors into four parts:
$f_{k,s}^{L(R)}= f_{k,s}^{L(R),W}+f_{k,s}^{L(R),\tilde{\chi}}+
             f_{k,s}^{L(R),f}+f_{k,s}^{L(R),\tilde{f}}$,
which correspond to loop diagrams of $W^{\pm}$, charged Higgs bosons
and ghost particles in Fig.1(b,4, b.5 and c.3), chargino loops in
Fig.1(c.2), quark and lepton loops in Fig.1(c.2), and scalar quark and lepton
loops in Fig.1(b.3, c.1)
respectively. Notations used in the appendix are defined below,
\begin{eqnarray*}
B_{mn}^{1 y} &=& B_{mn}[-p_1, m_{y}, m_{y}],~~~
B_{mn}^{2 y} = B_{mn}[-p_2, m_{y}, m_{y}],~~~
B_{mn}^{3 y} = B_{mn}[k_1+k_2, m_{y}, m_{y}],\\
C_{mn}^{y} &=& C_{mn}[-p_2, -p_1, m_{y}, m_{y}, m_{y}],\\
P_{B}&=&\frac{1}{\hat{s}-{m_{B}}^2+im_{B}\Gamma_{B}}~(B=h^0,H^0,A^0),~~~~
P_{G}=\frac{1}{\hat{s}-{m_{Z^0}}^2}.
\end{eqnarray*}
\par
The couplings of Higgs boson(B)-$\tilde{\chi}_j$-$\tilde{\chi}_i$
($\tilde{\chi}=\tilde{\chi}^{0}, \tilde{\chi}^{+}$) are denoted as
$$
V_{B\tilde{\chi}_{j}\tilde{\chi}_{i}}=
     V^{L}_{B\tilde{\chi}_{j}\tilde{\chi}_{i}}P_L+
     V^{R}_{B\tilde{\chi}_{j}\tilde{\chi}_{i}}P_R
$$
where $B=h^0,H^0,A^0,G^0$. All the explicit expressions of relevant
vertices can be found in the Appendix of \cite{hunter}. The form factors
are expressed explicitly as follows. $g$ represents electroweak coupling
constant; $Q_{f}$ denotes electrical charge($Q_{\tau}=-1$, $Q_{t}=2/3$, etc.)
and the factor 3 in $f_{k,s}^{f(\tilde{f})}$ arises from quark color.
\begin{eqnarray*}
f_{1,s}^{L,W}&=&
P_h \left\{ i [V_{h H^+ H^+} (B_{0}^{3 H^+}-4 C_{24}^{H^+})+
    V_{h G^+ G^+} (B_{0}^{3W}-{m_W}^2 C_{0}^{W}-4 C_{24}^{W})] \right.\\
&+& g {m_W} \left[ -\frac{B_{0}^{1W}+B_{0}^{2W}}{2}+ (\epsilon-3) B_{0}^{3W}+ ({m_W}^2
    +3 \hat{s}) C_{0}^{W}-\frac{\hat{s}}{2} (C_{11}^{W}+C_{12}^{W} \right.\\
&+& \left. \left. 2 C_{23}^{W})+ (16-5 \epsilon) C_{24}^{W} \right]
    \sin{(\alpha-\beta)} \right\} V_{h{\tilde\chi}^0_j {\tilde\chi}^0_i}^{L}\\
&+& (h^0 \to H^0,\sin{(\alpha-\beta)} \to -\cos{(\alpha-\beta)} ) ,\\
f_{13,s}^{L,W}&=&
8 i (P_h \left\{ V_{h H^+ H^+} (C_{12}^{H^+}+C_{23}^{H^+})+V_{h G^+ G^+} (
    C_{12}^{W}+C_{23}^{W}) \right.\\
&+& \left. i g {m_W} \left[ 2 C_{0}^{W}+3 (C_{12}^{W}+C_{23}^{W}) \right]
    \sin{(\alpha-\beta)} \right\} V_{h{\tilde\chi}^0_j {\tilde\chi}^0_i}^{L}\\
&+& (h^0 \to H^0,\sin{(\alpha-\beta)} \to -\cos{(\alpha-\beta)} ) ,\\
f_{1,s}^{L,\tilde{\chi}}&=&
- i \sum_{x=1}^{2}
    {m_{\tilde{\chi}_x^+}}(2 B_{0}^{1 \tilde{\chi}^+}-\hat{s} C_{0}^{\tilde{\chi}^+}-2 \hat{s}
    C_{12}^{\tilde{\chi}^+}-8 C_{24}^{\tilde{\chi}^+})
     \left[ P_h
    V_{h{\tilde\chi}^0_j {\tilde\chi}^0_i}^{L} (
    V_{h{\tilde\chi}^+_x {\tilde\chi}^-_x}^{L}+
    V_{h{\tilde\chi}^+_x {\tilde\chi}^-_x}^{R}) \right.\\
&+& \left. (h^0 \to H^0) +(h^0 \to A^0)+(h^0 \to G^0) \right] ,\\
f_{13,s}^{L,\tilde{\chi}}&=&
-4 i \sum_{x=1}^{2} \left[ C_{0}^{\tilde{\chi}^+}+4 (C_{12}^{\tilde{\chi}^+}+
    C_{23}^{\tilde{\chi}^+}) \right] {m_{\tilde{\chi}_x^+}} \left[ P_h
    V_{h{\tilde\chi}^0_j {\tilde\chi}^0_i}^{L} (
    V_{h{\tilde\chi}^+_x {\tilde\chi}^-_x}^{L}+
    V_{h{\tilde\chi}^+_x {\tilde\chi}^-_x}^{R}) \right.\\
&+& \left. (h^0 \to H^0) +(h^0 \to A^0)+(h^0 \to G^0) \right] ,\\
f_{21,s}^{L,\tilde{\chi}}&=&
4 \sum_{x=1}^{2} C_{0}^{\tilde{\chi}^+} {m_{\tilde{\chi}_x^+}} \left[ P_h
    V_{h{\tilde\chi}^0_j {\tilde\chi}^0_i}^{L} (
    V_{h{\tilde\chi}^+_x {\tilde\chi}^-_x}^{L}-
    V_{h{\tilde\chi}^+_x {\tilde\chi}^-_x}^{R})+
    (h^0 \to H^0) +(h^0 \to A^0) \right.\\
&+& \left. (h^0 \to G^0) \right] ,\\
f_{1,s}^{L,f}&=&
-2 i {Q_{\tau}}^2 {m_{\tau}} (2 B_{0}^{1\tau}-\hat{s} C_{0}^{\tau}-2 \hat{s} C_{12}^{\tau}-8 C_{24}^{\tau}) (P_h
    V_{h\tau\tau} V_{h{\tilde\chi}^0_j {\tilde\chi}^0_i}^{L}+P_H V_{H\tau\tau} V_{H{\tilde\chi}^0_j {\tilde\chi}^0_i}^{L})  \\
&+&  3(\tau \to t)+ 3(\tau \to b) ,\\
f_{13,s}^{L,f}&=&
-8 i {Q_{\tau}}^2 {m_{\tau}} \left[ C_{0}^{\tau}+4 (C_{12}^{\tau}+C_{23}^{\tau}) \right] (P_h
    V_{h\tau\tau} V_{h{\tilde\chi}^0_j {\tilde\chi}^0_i}^{L}+P_H V_{H\tau\tau} V_{H{\tilde\chi}^0_j {\tilde\chi}^0_i}^{L}) \\
&+&  3(\tau \to t)+ 3(\tau \to b) ,\\
f_{21,s}^{L,f}&=&
-8  {m_{\tau}} C_{0}^{\tau} (P_A V_{A\tau\tau}
    V_{A{\tilde\chi}^0_j {\tilde\chi}^0_i}^{L}+P_G V_{G\tau\tau}
    V_{G{\tilde\chi}^0_j {\tilde\chi}^0_i}^{L})+ 3(\tau \to t)+ 3(\tau \to b) ,\\
f_{1,s}^{L,\tilde{f}}&=&
  i {Q_{\tau}}^2 \sum_{x=1}^{2} (B_{0}^{3 \tilde{\tau}}-4 C_{24}^{\tilde{\tau}}) (P_h
    V_{h\tilde{\tau}_x\tilde{\tau}_x}
    V_{h{\tilde\chi}^0_j {\tilde\chi}^0_i}^{L}+P_H
    V_{H\tilde{\tau}_x\tilde{\tau}_x}
    V_{H{\tilde\chi}^0_j {\tilde\chi}^0_i}^{L})+ 3(\tau \to t)+ 3(\tau \to b) ,\\
f_{13,s}^{L,\tilde{f}}&=&
8 i {Q_{\tau}}^2 \sum_{x=1}^{2} (C_{12}^{\tilde{\tau}}+C_{23}^{\tilde{\tau}}) (P_h
    V_{h\tilde{\tau}_x\tilde{\tau}_x}
    V_{h{\tilde\chi}^0_j {\tilde\chi}^0_i}^{L}+P_H
    V_{H\tilde{\tau}_x\tilde{\tau}_x}
    V_{H{\tilde\chi}^0_j {\tilde\chi}^0_i}^{L})+ 3(\tau \to t)+ 3(\tau \to b),
\end{eqnarray*}
\begin{eqnarray*}
f_{i,s}^{R}=f_{i,s}^{L}(V_{B{\tilde\chi}^0_j {\tilde\chi}^0_i}^{L} \to
V_{B{\tilde\chi}^0_j {\tilde\chi}^0_i}^{R}),~~~(i=1,13,21~~B=h^0,H^0,A^0,G^0).
\end{eqnarray*}
\par
In the above expressions we adopted the definitions of one-loop integral
functions in \cite{s13} and defined $d=4-2 \epsilon$. The numerical
calculation of the vector and tensor loop integral functions can be
traced back to four scalar loop integrals $A_{0}$, $B_{0}$, $C_{0}$,
$D_{0}$ as shown in \cite{s14}.

\vskip 10mm
\begin{flushleft} {\bf Figure Captions} \end{flushleft}

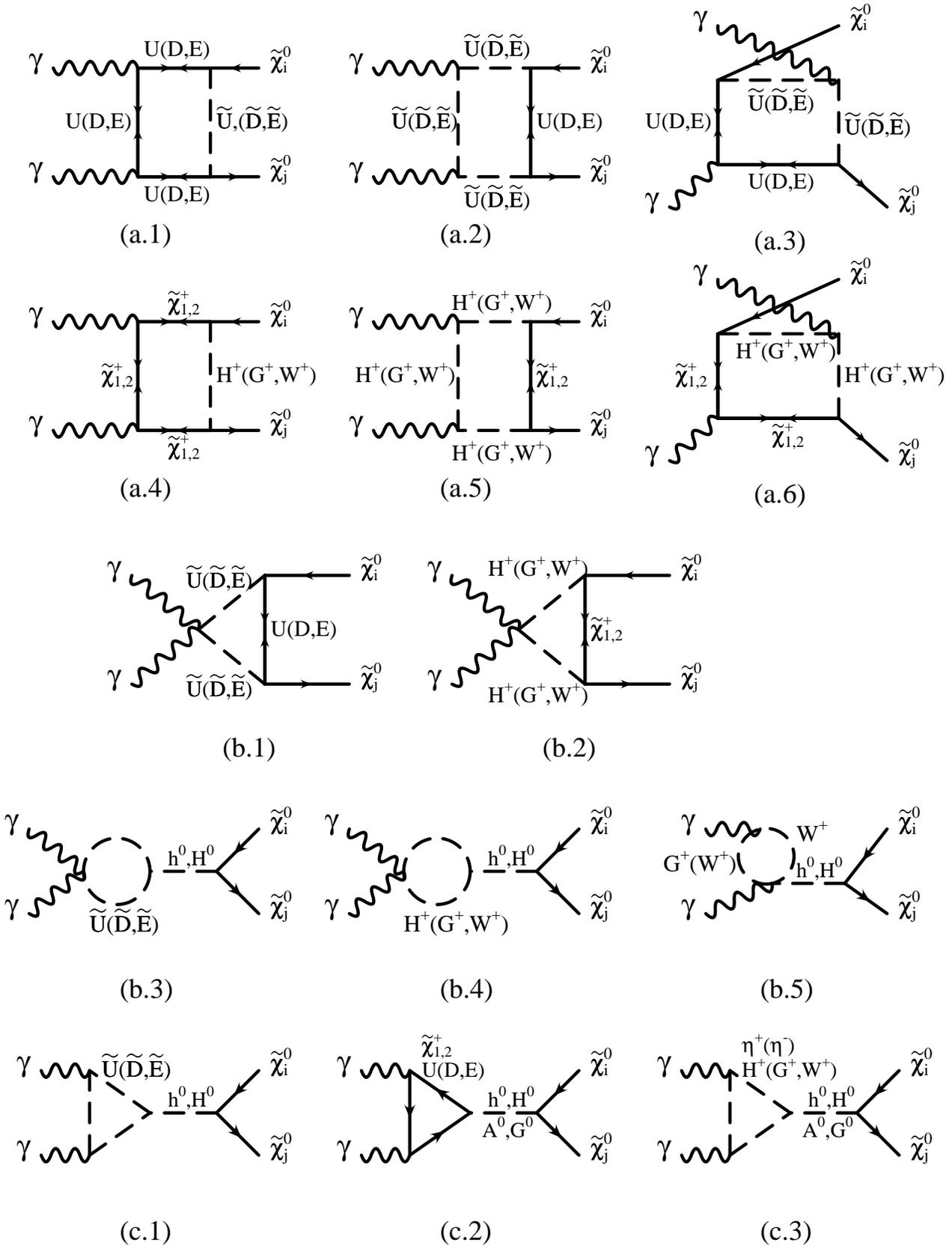
{\bf Fig.1} The Feynman diagrams of the subprocess $\gamma \gamma \rightarrow
\tilde{\chi}^{0}_{i}\tilde{\chi}^{0}_{j}$. ($a.1 \sim 6$) Box diagrams.
($b.1 \sim 5$) Quartic interaction diagrams. ($c.1 \sim 3$) Triangle diagrams.

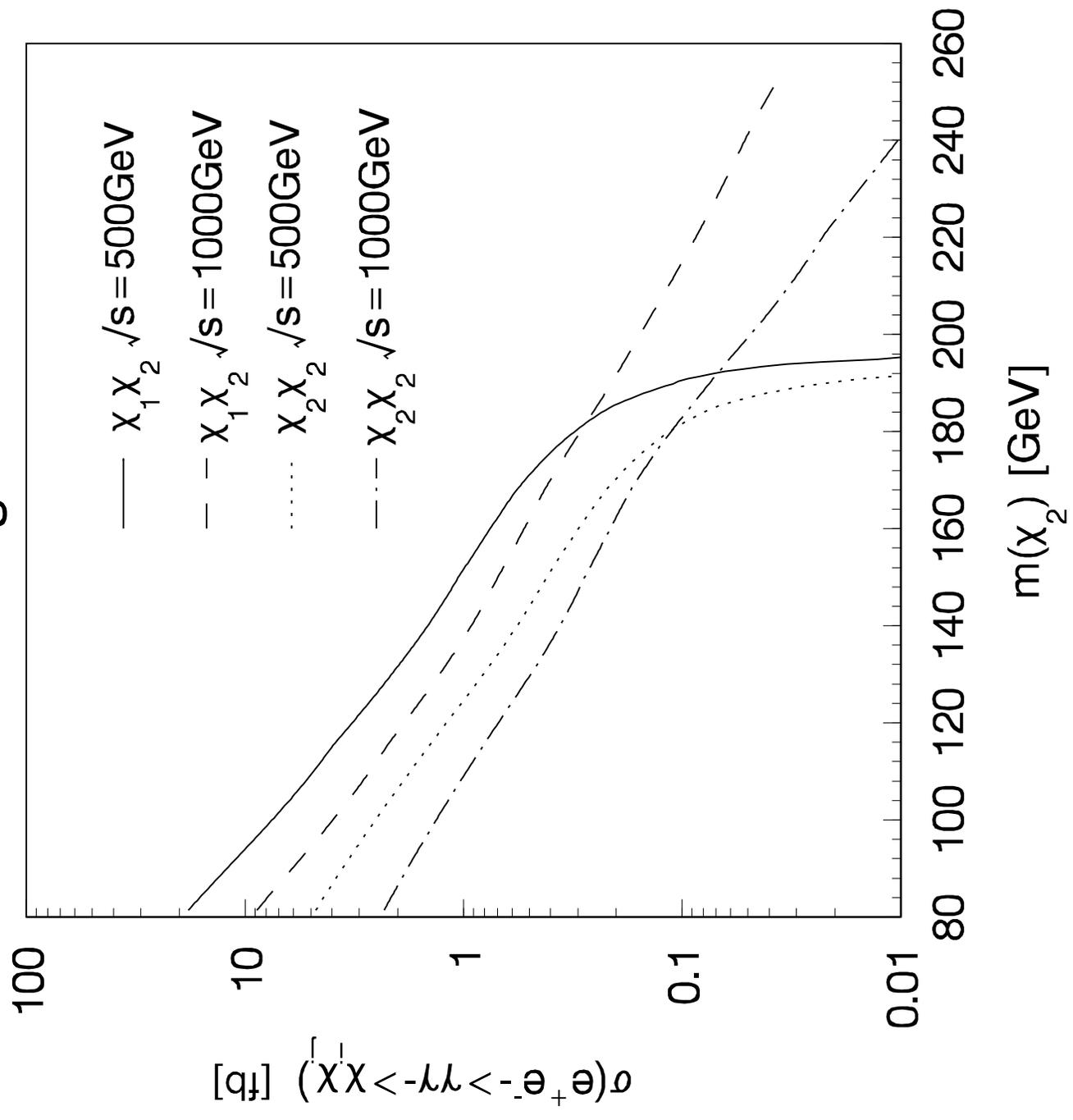
{\bf Fig.2} Total cross sections of the process $e^{+} e^{-} \rightarrow
\gamma \gamma \rightarrow \tilde{\chi}^{0}_{i}\tilde{\chi}^{0}_{j},~~
(i=1,2,~j=2)$ as function of $m_{\tilde{\chi}1/2}$ at electron-positron
LC with $\sqrt{s}=500~GeV$. The solid curve and dashed curve are for
$\tilde{\chi}^{0}_{1}\tilde{\chi}^{0}_{2}$ production at LC with
$\sqrt{s}=500~GeV,1000~GeV$, respectively. The dotted curve and dash-dotted
curve are for $\tilde{\chi}^{0}_{2}\tilde{\chi}^{0}_{2}$ production at
LC with $\sqrt{s}=500~GeV$ and $1000~GeV$, respectively.

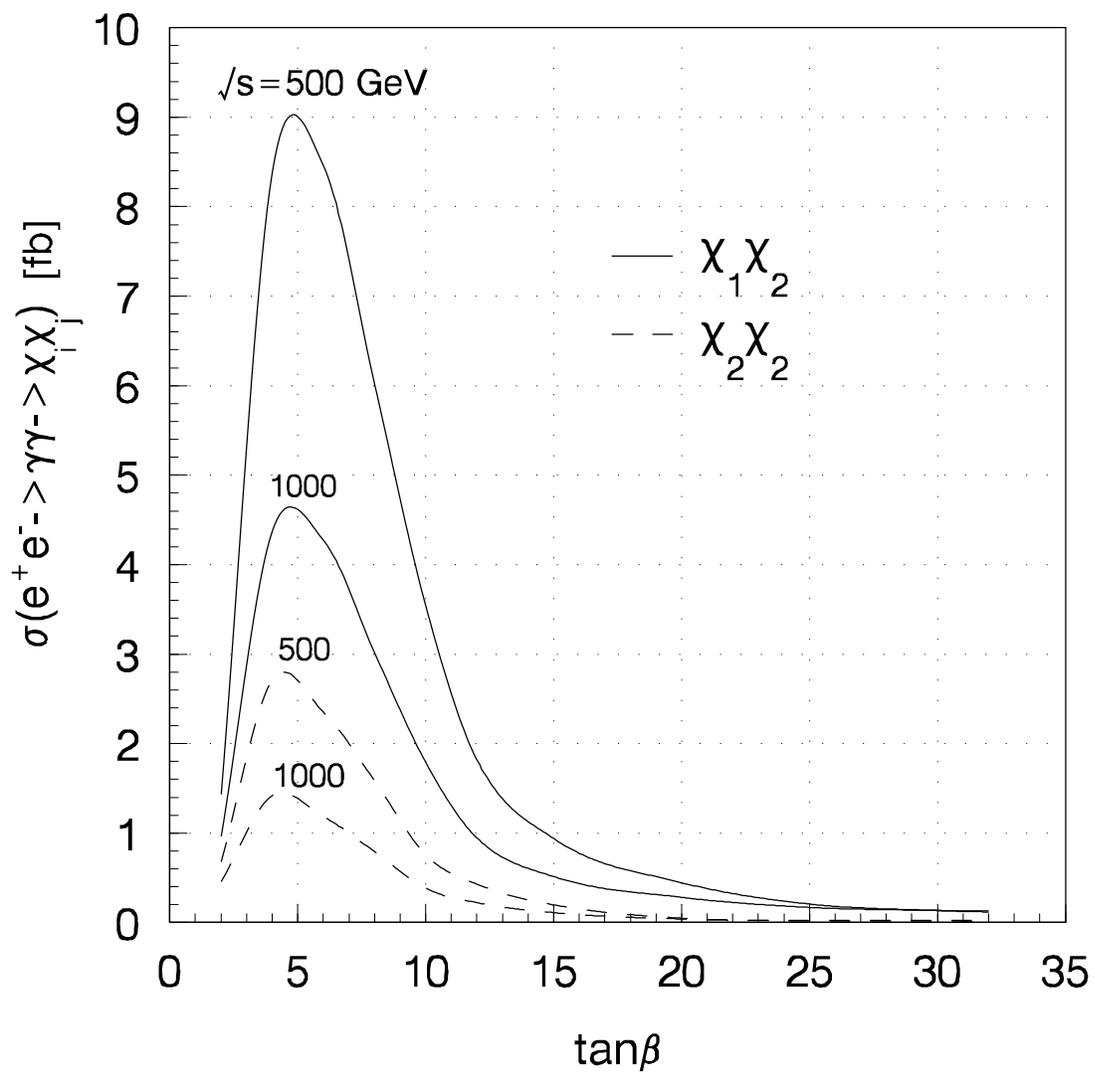
{\bf Fig.3} Total cross sections of the process $e^{+} e^{-} \rightarrow
\gamma \gamma \rightarrow \tilde{\chi}^{0}_{i}\tilde{\chi}^{0}_{j},~~
(i=1,2,~j=2)$ as function of $\tan\beta$. The dashed-lines are for
$\tilde{\chi}^{0}_{2} \tilde{\chi}^{0}_{2}$ production and the
full-line curves are for $\tilde{\chi}^{0}_{1}\tilde{\chi}^{0}_{2}$
production, with $e^+e^-$ colliding energy $\sqrt{s}=500~GeV$ and $1000~GeV$
respectively.

\end{large}
\end{document}